\documentclass[12pt,preprint,showpacs,preprintnumbers]{revtex4}
\usepackage{amssymb}
\usepackage{amsmath}
\usepackage{graphicx}
\usepackage{dcolumn}
\usepackage{bm}
\usepackage{epsfig}
\usepackage[T1]{fontenc}
\usepackage{ae,aecompl}
\usepackage{subeqnarray}
\usepackage{cases}
\begin{document}
\graphicspath{{figures/}}
\title{Lattice Boltzmann model for combustion and detonation}
\author{Bo Yan,
Aiguo Xu\footnote{%
Corresponding author. Email address: Xu\_Aiguo@iapcm.ac.cn}
Guangcai Zhang\footnote{%
Corresponding author. Email address: Zhang\_Guangcai@iapcm.ac.cn},
Yangjun Ying,
Hua Li
}
\affiliation{
National Key Laboratory of Computational Physics, \\
Institute of Applied Physics and Computational Mathematics, P. O. Box 8009-26, Beijing 100088, P.R.China
}
\date{\today }

\begin{abstract}
In this paper we present a lattice Boltzmann model for combustion and detonation. In this model the fluid behavior is described by a finite-difference lattice Boltzmann model by Gan, Xu, Zhang, et al [Physica A 387 (2008) 1721]. The chemical reaction is described by the Lee-Tarver model [Phys. Fluids 23 (1980) 2362]. The reaction heat is naturally coupled with the flow behavior. Due to the separation of time scales in the chemical and thermodynamic processes, a key technique for a successful simulation is to use the operator-splitting scheme. The new model is verified and validated by well-known benchmark tests. As a specific application of the new model, we studied the simple steady detonation phenomenon. To show the merit of LB model over the traditional ones, we focus on the reaction zone to study the non-equilibrium effects. It is interesting to find that, at the von Neumann peak, the system is nearly in its thermodynamic equilibrium. At the two sides of the von Neumann peak, the system deviates from its equilibrium in opposite directions.
In the front of von Neumann peak, due to the strong compression from the reaction product behind the von Neumann peak, the system experiences a sudden deviation from thermodynamic equilibrium. Behind the von Neumann peak, the release of chemical energy results in thermal expansion of the matter within the reaction zone, which drives the system to deviate the thermodynamic equilibrium in the opposite direction.  From the deviation from thermodynamic equilibrium, $\boldsymbol{\Delta}_m^*$, defined in this paper, one can understand more on the macroscopic effects of the system due to deviating from its thermodynamic equilibrium.

\end{abstract}

 \maketitle

\section{Introduction}

In recent two decades the Lattice Boltzmann (LB) method has been becoming a powerful tool for simulating complex systems \cite{succi,benzi,xu1,xu2,aidun}. Unlike traditional computational fluid dynamics methods, the fundamental idea of LB method is to construct simplified kinetic models that incorporate the essential physics of microscopic or mesoscopic processes. In the continuum limit the LB results should obey the macroscopic equation such as Euler equation and Navier-Stokes equation. Because of its intrinsic kinetic nature, the LB model can be used to investigate many physical phenomenon at microscopic or mesoscopic level which are generally difficult for traditional methods. It has been successfully used to study various complex fluids such as magnetohydrodynamics \cite{chen-mar,vahala-keat}, flows of suspensions \cite{ladd}, flows through porous media \cite{succi-porous,li-porous}, compressible fluid dynamics \cite{watari,xu2005,gan-compres,li-compres}, multiphase flows \cite{phase1,phase2,phase3,phase4,phase-gan1,phase-gan2,front2012}, etc. It should be pointed out that most of these studies are focused on isothermal and incompressible systems. Given the importance of shock wave and detonation in science and engineering, LB simulation of high speed compressible fluids were attempted from the early days of LB research. But most of current LB models for compressible flows are still subject to the constraint of low Mach number. Recently, high speed compressible LB model is constructed from the following few aspects \cite{front}. (i) Consider appropriate additional viscosity for the situation where the system is far from equilibrium, for example, for systems with shocks \cite{gan-compres,phase-gan1,vis-pan,vis-chen,vis-gorban}. (ii) Construct Multiple-Relaxation-Time (MRT) LB model \cite{succiMRT,luoMRT,chenMRT}. (iii) Establish entropic LB model \cite{entro-succi,entro-karlin}. (iv) Introduce the FIX-UP \cite{entro-succi,fixup-li} and/or flux-limiter schemes \cite{phase3,GanFlux,ChenFlux}. In this work our LB model for high speed compressible flows belongs to the first category.

Detonation \cite{fickett} is the supersonic combustion that usually propagates through shock wave. It is different from deflagration which is the subsonic combustion usually propagating via thermal conductivity. The detonation phenomena are highly concerned issues in science and engineering. They are widely used in the acceleration of various projectiles, mining technologies, depositing of coating to a surface or cleaning of equipment, etc. However, unintentional detonation phenomenon, when deflagration is desired, is a problem in some devices. The detonation phenomenon was first recognized by Berthelot et al \cite{Berthelot} and Mallard et al \cite{Mallard} during experiments of flame propagation. The Chapmann-Jouguet (CJ) theory \cite{C-J} at alternation of the 20 century considers that the detonation front can be treated as a strong discontinuity plane with chemical reaction and that the chemical reaction immediately completes within the infinite thin region. The Zeldovich-von Neumann-Doering (ZND) model \cite{Zeld,Neumann,Doering} presented in 1940s treats detonation front as a shock wave in which no chemical reaction occurs. In this model the chemical reaction is triggered by shock wave and proceeds at a finite rate to completion thereafter.

Numerical simulations for combustion and detonation have been significantly improved in recent 30 years along with the development in both computational methods and available computer facilities\cite{Mader,WangC1,WangC2}. There are still two main challenges in simulating such processes. The first is how to describe wave front. Traditional methods include the Eulerian cell scheme \cite{euler1,euler2,euler3,euler4}, Arbitrary-Lagrangian-Eulerian(ALE) algorithm \cite{ale1,ale2}, level set method \cite{levelset1,levelset2},  Volume of Fluid(VOF) method \cite{vof}, front tracking method \cite{tracking1,tracking2,tracking3}, etc. Though the traditional methods can treat contact discontinuity surface, they are not capable of describing the detonation wave faithfully. The second is how to simulate the chemical reaction and energy releasing processes which are coupled with the flow behavior. Several chemical reaction models \cite{Sun-Zhu}, such as the Arrhenius kinetics, Cochran's rate function, forest fire burn, two-step model, Lee-Tarver model, etc, have been introduced. During the energy releasing procedure, the state of system is generally far from equilibrium. However, the traditional description based on computational fluid dynamics considers only the cases where the systems can be described by the continuum Euler or Navier-Stokes equations. As a special discretization of the basic equation of non-equilibrium statistical physics, the Boltzmann equation, the LB model is intrinsically suitable for handling  the non-equilibrium effects of the problems mentioned above.

To the best of our knowledge, The earliest work of LB modeling combustion phenomena was finished by Succi, et al \cite{succi-comb}. They simulated the methane-air laminar flame under the assumptions of fast chemistry and cold flames with weak heat release. In 2000, Filippova et al \cite{filippova} proposed a hybrid scheme where the flow field is solved by LB method, while the transport equation, energy equation and components equation are solved by the finite difference method. As a incomplete LB scheme, the hybrid scheme does not apply fully the advantage of LB method. Later, Yamamoto et al \cite{yamamoto} employed a pure LB scheme for combustion simulation. The scheme uses double distribution functions. One is employed to describe the density and velocity fields, the other describes temperature field. Lee et al \cite{lee} proposed a quasi-incompressible model for the laminar jet diffusion flame. All those LB models mentioned above are for nearly incompressible systems which are not appropriate for detonation phenomena. In addition to incompressibility, they assumed that the chemical reaction does not affect the flow fields, which is a lethal flaw for simulating detonation and most of combustion problems.

In this paper, a novel model for combustion and detonation is proposed. It couples the Finite-Difference (FD) LB model by Gan, Xu, Zhang et al \cite{phase-gan1} for fluid with the Lee-Tarver \cite{lee-tarver} model for chemical reaction. The Lee-Tarver reaction rate function is composed of two terms, the first is generally referred to as the hot spot formation term and the second is generally referred to the reaction growth term. The former is used to investigate various spot formation processes and their subsequent growth. The latter describes the growth of the reaction. The Lee-Tarver reaction model is one of the most physically justifiable models which have produced satisfying simulations of the experimental results. The rest part of the paper is organized as below. In section II we introduce the FDLB model for combustion and detonation.  In Section III we validate the new model by simulating various detonation problems. In section IV, we define explicitly some quantities to measure the deviation from thermodynamic equilibrium from various aspects and use the new LB model to study the non-equilibrium effects in the steady detonation procedure.
Conclusions and discussions are given in section V.

\section{FDLB model for combustion and detonation}

In this section we construct a hybrid LB model for simulating the combustion and detonation phenomena. Such a model should be capable of describing the high speed compressible flow influenced by the chemical reaction. For the first step, it is assumed that there are only two species present, the reactant and the product, and the reactant is converted to the product by a one step irreversible chemical reaction governed chemical kinetics equation.
The flow field is described by a FDLB model.

\subsection{FDLB model for flow behavior}

For modeling the flow behavior in the combustion and detonation process, we employ the lattice Bhatnagar-Gross-Krook (BGK) model improved by Gan, Xu, Zhang, et al \cite{phase-gan1} for the density, momentum and energy. The model is composed of three components: the discrete velocity model, modified Lax-Wendroff finite difference scheme and additional viscosity term.

The discrete velocity model has the following 33 discrete velocities
\begin{subequations}
\label{dvm}
\begin{numcases}{}
\textbf{v}_0=0,\\[2mm]
\textbf{v}_{ki}=v_k\Big{[}\cos\Big{(}\frac{i\pi}{4}\Big{)},
\sin\Big{(}\frac{i\pi}{4}\Big{)}\Big{]},~~i=1,2\ldots8 \text{,}
\end{numcases}
\end{subequations}
where subscript $k$ and $i$ indicate the $k$-th group, the $i$-th direction of the particle velocities, respectively. For the compressible fluid, the evolution of distribution function $f_{ki}$ with BGK approximation reads
\begin{equation}
\frac{\partial f_{ki}}{\partial t}+\textbf{v}_{ki}\cdot\frac{\partial f_{ki}}{\partial\textbf{r}}=
-\frac{1}{\tau}[f_{ki}-f^{eq}_{ki}],
\label{lbe}
\end{equation}
where $\textbf{r}$ and $\tau$ are spatial coordinate and the relaxation time. The equilibrium distribution function $f_{ki}^{eq}$ is calculated by
\begin{eqnarray}
f^{eq}_{ki}&=&\rho F_k\Big{[}\Big{(}1-\frac{u^2}{2T}+\frac{u^4}{8T^2}\Big{)}+
\frac{v_{ki\epsilon}u_\epsilon}{T}\Big{(}1-\frac{u^2}{2T}\Big{)}+
\frac{v_{ki\epsilon}v_{ki\pi}u_\epsilon u_\pi}{2T^2}\Big{(}1-\frac{u^2}{2T}\Big{)}\nonumber\\[5mm]
&~&+\frac{v_{ki\epsilon}v_{ki\pi}v_{ki\vartheta}u_\epsilon u_\pi u_\vartheta}{6T^3}+
\frac{v_{ki\epsilon}v_{ki\pi}v_{ki\vartheta}v_{ki\xi}u_\epsilon u_\pi u_\vartheta u_\xi}{24T^4}\Big{]},
\end{eqnarray}
where $\rho$ and $T$ are density and temperature respectively. Weight factors $F_k$ and $F_0$ are calculated by
\begin{eqnarray*}
F_k &=& \frac{1}{v^2_k(v^2_k-v^2_{k+1})(v^2_k-v^2_{k+2})(v^2_k-v^2_{k+3})}\Big{[}48T^4-
6(v^2_{k+1}+v^2_{k+2}+v^2_{k+3})T^3\\[5mm]
&~&+(v^2_{k+1}v^2_{k+2}+v^2_{k+2}v^2_{k+3}+v^2_{k+3}v^2_{k+1})T^2-\frac{v^2_{k+1}v^2_{k+2}v^2_{k+3}}{4}T\Big{]},~~k=1,2,3,4\\
F_0&=&1-8(F_1+F_2+F_3+F_4),
\end{eqnarray*}
where
\begin{equation}
\{k+l\}=
\begin{cases}k+l, & \text{if}~~ k+l\leq4,\nonumber\\
k+l-4, & \text{if}~~k+l>4.
\end{cases}
\end{equation}
We choose $v_0=0$ and four nonzero $v_k$ $(k=1,2,3,4)$.

The following moments of the equilibrium distribution function are necessary for the current FDLB model:
\begin{eqnarray}
\label{mo1}&~&\sum_{ki}f^{eq}_{ki}=\rho,\\
\label{mo2}&~&\sum_{ki}v_{ki\alpha}f^{eq}_{ki}=\rho u_\alpha,\\
\label{mo3}&~&\sum_{ki}\frac{1}{2}v_{ki}^2f^{eq}_{ki}=e_{therm}+\frac{1}{2}\rho u^2=\rho T+\frac{1}{2}\rho u^2=P+\frac{1}{2}\rho u^2,\\
\label{mo4}&~&\sum_{ki}v_{ki\alpha}v_{ki\beta}f^{eq}_{ki}=e_{therm}\delta_{\alpha\beta}+\rho u_\alpha u_\beta,\\
\label{mo5}&~&\sum_{ki}v_{ki\alpha}v_{ki\beta}v_{ki\gamma}f^{eq}_{ki}=e_{therm}(u_\gamma\delta_{\alpha\beta}
+u_\alpha\delta_{\beta\gamma}+u_\beta\delta_{\gamma\alpha})+\rho u_\alpha u_\beta u_\gamma,\\
\label{mo6}&~&\sum_{ki}\frac{1}{2}v_k^2v_{ki\alpha}f^{eq}_{ki}=2e_{therm}u_\alpha+\frac{1}{2}\rho u^2u_\alpha,\\
\label{mo7}&~&\sum_{ki}\frac{1}{2}v_k^2v_{ki\alpha}v_{ki\beta}f^{eq}_{ki}=
\Big{[}2T+\frac{1}{2}u^2\Big{]}e_{therm}\delta_{\alpha\beta}+
\Big{[}3e_{therm}+\frac{1}{2}\rho u^2\Big{]}u_\alpha u_\beta,
\end{eqnarray}
where $\textbf{u}$, $e_{therm}$, $P$ are the hydrodynamic velocity, internal energy and pressure, respectively. We can obtain $\rho$, $\textbf{u}$, $T$ and $P$ from equations \eqref{mo1} to \eqref{mo3}.

Performing the Lax-Wendroff finite difference scheme to the left hand side of equation \eqref{lbe}, adding the dispersion term and additional viscosity to the right, then using second-order space-centered difference to the right hand of equation \eqref{lbe}, we eventually obtain the following LB equation,
\begin{eqnarray}
f^{new}_{kiI}&=&f_{kiI}-\frac{c_{ki\alpha}}{2}(f_{kiI+1}-f_{kiI-1})-\frac{\Delta t}{\tau}[f_{kiI}-f_{kiI}^{eq}]\nonumber\\
&~&+\frac{c^2_{ki\alpha}}{2}(f_{kiI+1}-2f_{kiI}+f_{kiI-1})\nonumber\\
&~&+\frac{c_{ki\alpha}(1-c^2_{ki\alpha})}{12}(f_{kiI+2}-2f_{kiI+1}+2f_{kiI-1}-f_{kiI-2})\nonumber\\
&~&+\frac{\theta_{\alpha I}|\kappa_\alpha|(1-|\kappa_\alpha|)}{2}(f_{kiI+1}-2f_{kiI}+f_{kiI-1}),
\label{laxbgk}
\end{eqnarray}
where
\begin{eqnarray*}
c_{ki\alpha}&=&v_{ki\alpha}\Delta t/\Delta r_\alpha,~~\kappa_\alpha=u_\alpha\Delta t/\Delta r_\alpha,\\[5mm]
\theta_{\alpha I}&=&\eta\Bigg{|}\frac{P_{\alpha I+1}-2P_{\alpha I}+P_{\alpha I-1}}
{P_{\alpha I+1}+2P_{\alpha I}+P_{\alpha I-1}}\Bigg{|}.
\end{eqnarray*}
The third suffixes, $I-2$, $I-1$, $I$, $I+1$, $I+2$, index the meshs node in $x$- or $y$-direction.
As a monotonicity preserving scheme, the Lax-Wendroff scheme improves the numerical stability. The fourth line of equation \eqref{laxbgk} is the additional viscosity which makes the scheme suitable for both the low and high speed fluid flows.

The FDLB scheme gives consistent results with the Naiver-Stokes equations in the continuum limit. The traditional numerical simulation for combustion and detonation is to solve the governing equations which couple the hydrodynamic equations with the chemical kinetic equation. In most of traditional studies, the effects of transportation due to viscosity and heat conduction are not explicitly taken into account,  so what used are the simple Euler equations. Currently, the combustion and detonation behaviors with viscosity and heat conduction still need to be carefully investigated. The LB model contains more fundamental information than the traditional description, especially that related to non-equilibrium effects. This characteristics will be further discussed and illustrated below.

\subsection{Lee-Tarver model for combustion}

Selecting appropriate chemical reaction kinetics is an  important step for describing  the combustion and detonation phenomena under consideration. The reaction model decides the initiation, the development and the propagation of the combustion/detonation in the simulation. The chemical reaction process is very complex. It includes variety of reaction mechanism. So far, most of the chemical reaction kinetics are phenomenological models. The Lee-Tarver model \cite{lee-tarver} has produced many satisfying simulations of the experimental results. It is widely used in combustion and detonation studies. The Lee-Tarver chemical reaction rate law reads
\begin{eqnarray}
\label{leetarver1}\frac{d\lambda}{dt}&=&a(1-\lambda)^x\eta^r+b(1-\lambda)^x\lambda^yP^z,\\[3mm]
\eta&=&\frac{V_0}{V}-1.
\label{leetarver2}
\end{eqnarray}
Here $\lambda$ is parameter for the chemical reaction process or the fraction of reactant that has reacted. $\eta$ is the relative compression. $V_0$ and $V$ are specific volume ahead of the shock front and behind it, respectively. $a$, $b$, $x$, $y$, $z$, $r$ are constant parameters, where $b$, $x$, $y$ describe the dependence on the variables like burning area, etc. The constant $b$ is also dependent on the pressure of incident shock wave. $z$ describes the dependence on the local pressure. This model is composed of two terms: The first term in equation \eqref{leetarver1} is generally referred to as the hot spot formation term, which mainly describes the formation and subsequent growth of the hot spots. The second term in equation \eqref{leetarver1} is generally referred to as the growth term. It mainly describes the growth of the reaction.

As the first step, in this work we consider only the simplest case where $x=y=1$, $z=r=0$. Considering the thermal initiation, the Lee-Tarver model and initiation condition are written in the following form
\begin{equation}
\label{lt}
\frac{d\lambda}{dt}=
\begin{cases}a(1-\lambda)+b(1-\lambda)\lambda, & T \geq T_{th}~\text{and}~ 0 \leq \lambda \leq 1,\\
0, & \text{else},
\end{cases}
\end{equation}
where $T_{th}$ is the temperature threshold for chemical reaction.

Since the chemical reaction is much faster than the process of fluid flow, the time scale of chemical reaction and macroscopic behavior differ by several orders, the term of chemical reaction can not be treated using the same time step as the terms describing fluid flow. In this paper, we apply the operator-splitting scheme to equation \eqref{lt}. In numerical simulations the evolution can be described by two steps:

Step 1, calculate the convection contribution
 \begin{equation}
\label{convection}
\frac{\partial\lambda}{\partial t}+u\nabla\lambda=0.
\end{equation}

Step 2, calculate the contribution of chemical reaction
 \begin{equation}
\label{diffusion}
\frac{\partial\lambda}{\partial t}=a(1-\lambda)+b(1-\lambda)\lambda.
\end{equation}

Equation \eqref{convection} can be solved by the upwind scheme
\begin{equation} \frac{\lambda^{n+1}_{I}-\lambda^n_{I}}{\Delta t}=-
\begin{cases} \frac{u(\lambda^n_{I}-\lambda_{I-1}^n)}{\Delta x} & u\geq0, \\
\frac{u(\lambda_{I+1}^n-\lambda_{I}^n)}{\Delta x} & u<0,
\end{cases}
\end{equation}
the suffixes $I-1$, $I$, $I+1$ index the mesh node of in $x$- or $y$-direction. Equation \eqref{diffusion} can be solved analytically. It gives
 \begin{equation}
\label{exact}
\lambda_I^{n+1}=\frac{e^{(a+b)\Delta t}+\frac{a(\lambda_I^n-1)}{a+b\lambda_I^n}}{e^{(a+b)\Delta t}+\frac{b(1-\lambda_I^n)}{a+b\lambda_I^n}}.
\end{equation}

\subsection{Coupling of the reaction and flow behavior}

In previous LB studies \cite{succi-comb,filippova,yamamoto,lee}, it was assumed that the chemical reaction does not affect the flow fields, which is a lethal flaw for simulating detonation and most of combustion problems. In real combustion process, the heat of reaction is coupled with the internal energy. The increment of internal energy results in the variation of pressure which naturally influences the flow behavior. Accordingly, in our LB simulation  the total internal energy increasing rate $\dot{e}$ contains two parts: the chemical reaction part $\dot{e}_{chem}$ and the original thermodynamic part $\dot{e}_{therm}$. It can be written as below:
\begin{eqnarray}
\dot{e}=\dot{e}_{therm}+\dot{e}_{chem},\label{totel} \\
\dot{e}_{{chem}}=\dot{\lambda}\rho Q\label{echem} ,
\end{eqnarray}
where $Q$ is the reaction heat per unit mass of reactant.

In our simulations, we first obtain $e_{therm}$  from equation \eqref{mo3}. Then calculate the total internal energy $e$ via equations \eqref{totel} and  \eqref{echem}. Nextly, update the pressure $P$ and temperature $T$ by replacing  $e_{therm}$ by the total internal energy $e$. Finally, calculate equilibrium distribution function $f_{ki}^{eq}$ by using the updated temperature $T$. In this way, the chemical reaction couples naturally with the flow behavior.

\section{Verification and validation}

In this section, we study several typical one- and two-dimensional detonation problems using the present model: the piston problem including effects of viscosity and heat conduction which is a typical test in studies on shock initiation of explosive; collision between detonation and shock waves; regular and Mach reflection of plane detonation wave; Richtmyer-Meshkov instability caused by detonation wave.

\subsection{Piston problem with viscosity and heat conduction}

Consider a detonation wave in rigid tube, followed by a piston controlled by external forces so that its velocity may be specified. Three cases are considered: (i) $u_p >u_{cj}$, where $u_p$ is the velocity of piston, $u_{cj}$ is velocity of CJ point. This case has a very simple solution: a uniform state between the front and the piston, since the flow is subsonic, a rarefaction generated at the piston will overtake the front and eventually produce the uniform steady solution corresponding to the new piston velocity; (ii) $u_p =u_{cj}$. We still obtain a uniform state, the CJ state. As the flow is sonic, the rarefaction generated at the piston can not overtake the detonation wave front; (iii) $u_p <u_{cj}$. The front can move no slower than velocity of detonation $D_{cj}$, so we still have the CJ state at the front, but the rarefaction wave reduces the velocity behind the front gradually to that of the piston. Therefore, the solution consists of a rarefaction wave followed by a uniform state.

\begin{figure}[!h]
\begin{center}
\includegraphics[width=6cm]{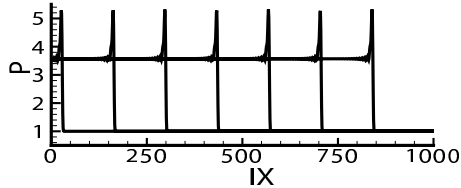}
\caption{\label{piston} Pressure profiles of case (ii) of the piston problem including effects of viscosity and heat conduction at times $0$, $0.05$, $0.1$, $0.15$, $0.2$, $0.25$ and $0.3$. ``IX" is the index of lattice node in the $x$-direction.}
\end{center}
\end{figure}

In order to demonstrate the validity of the new model, we numerical simulate the case (ii) of piston problem, see figure \ref{piston}. The initial macroscopic quantities are set as below
\begin{subequations}
\label{pistoncase2}
\begin{numcases}{}
(\rho,u,v,T,\lambda)_L=(1.34531,0.809765,0,2.65175,1),\\[5mm]
(\rho,u,v,T,\lambda)_R=(1,0,0,1,0).
\end{numcases}
\end{subequations}
The periodic boundary condition is adopted on the top and bottom. The left boundary is wall, and at the right boundary we impose the right states of the initial traveling wave solution. Other parameters are $ \Delta x=\Delta y=0.001$, $\Delta t =10^{-5}$, $\tau=8\times10^{-6}$, $a=1.0$, $b=10^3$, $Q=1.0$, $\gamma=2.0$, $T_{th}=1.1$ and $N_x \times N_y =1000 \times 3$, where $ix$ and $iy$ are the indexes of lattice node in the $x$- and $y$-directions, $N_x$ and $N_y$ are the numbers of lattice node in the $x$- and $y$-direction.

Figure \ref{piston} shows the pressure profile of case (ii) $(u_p =u_{cj})$ of piston problem include effect of viscosity and heat conduction at time $0, 0.05, 0.1, 0.15, 0.2, 0.25, 0.3$. The essential features of detonation, such as the von Neumann peak, the chemical reaction zone, etc can be observed obviously.

The Hugoniot relations of detonation wave read

\begin{equation}
\label{hugoniot1}\rho_0(D-u_0)=\rho_1(D-u_1),
\end{equation}
\begin{equation}
\label{hugoniot2}p_1-p_0=\rho_0 (D-u_0)(u_1-u_0),
\end{equation}
\begin{equation}
\label{hugoniot3}e_1-e_0=\frac{1}{2}(p_1+p_0)(1/\rho_0-1/\rho_1)+\lambda Q,
\end{equation}
where the suffixes $0$ and $1$ index physical quantities before and after the detonation wave, respectively, $D$ is the velocity of detonation wave. Removing $\lambda Q$, \eqref{hugoniot1}-\eqref{hugoniot3} reduces to the Hugoniot relations of shock wave. From figure \ref{piston}, we obtain the velocity of detonation wave $D=3.1$. Other physical quantities after detonation wave are $(\rho,u,p)=(1.34587,0.81226,3.54092)$ from current model. To numerically validate the current model, we apply the above LB results to \eqref{hugoniot1}-\eqref{hugoniot3}. The values of the Left Hand Side(LHS) are $3.1$, $2.54092$, $1.63095$ and values of the Right Hand Side(RHS) are $3.079$, $2.51801$, $1.58348$.
Compared with the values of LHS, the relative differences of the two sides are about $0.7\%$, $0.9\%$ and $3\%$, respectively. Considering that the viscosity and heat conductivity are not taken into account in the Hugoniot relations \eqref{hugoniot1}-\eqref{hugoniot3}, the LB results are satisfying. Simulation results fully indicate that the LB model works for simulating the shock initiation of explosives.

\subsection{Collision between detonation and shock waves}

The setup is as follows: a detonation (on the left side) and a shock wave (on the right side) move face to face in the opposite directions. In order to investigate the collision between detonation and shock waves, we assume the shock strength can't ignite the combustible gas, that is, the temperature after the shock is lower than temperature threshold of explosion.  The physical and chemical processes after collision can be divided into several stages: Firstly, the collision between detonation and shock is instantaneous. Then, on the left hand side of the contact discontinuity resulted from the difference of entropies, the transmitted shock interacts with combustion production. On the right hand side of the contact discontinuity, the transmitted detonation interacts with the combustible gas compressed by shock and consequently chemical reaction occurs. The zone between transmitted shock and transmitted detonation is the domain of influence. To simulate the interaction of detonation wave and shock wave, the initial strength of the shock wave must be weaker than that of the detonation wave.

\begin{figure}[!h]
\begin{center}
\includegraphics[width=10cm]{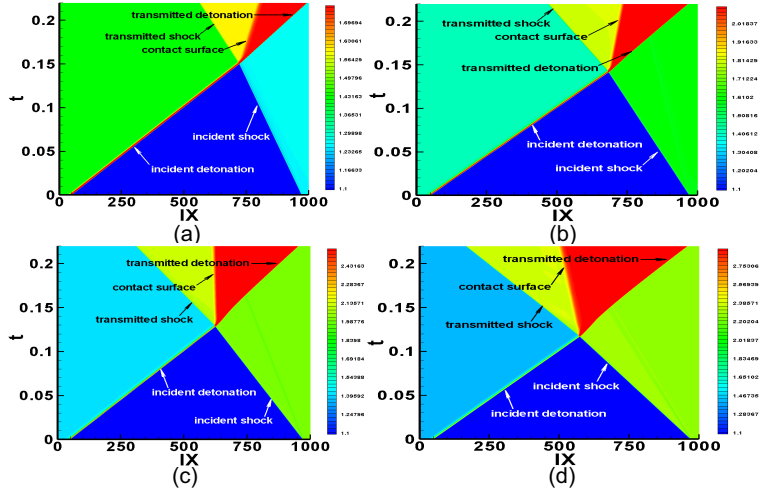}
\caption{\label{collision} Density distribution on the $x$-$t$ plane of detonation/shock collisions with four different shock strengths. (a) $M_s=1.2$. (b) $M_s=1.5$. (c) $M_s=2.0$. (d) $M_s=2.5$.}
\end{center}
\end{figure}

Density distributions on the $x$-$t$ plane for the procedure of detonation/shock collision with different shock strengths are shown in figures \ref{collision}(a)-(d). The initial macroscopic quantities on the left boundary are $\rho_L =1.41995$, $u_L =1.52086$, $v_L =0$, $T_L =6.11682$, $\lambda_L =1.0$. The parameters are $Q=3.5$, $T_{th}=3.8$. Other parameters are chosen as the same as figure \ref{piston}. Thus, the Mach number $M_D$ of the detonation wave is about $3.5$. The Mach numbers for shock waves on the right side are $M_s=1.2$, $1.5$, $2.0$, and $2.5$, respectively. Incident shock, incident detonation, transmitted shock, transmitted detonation and contact surface are obtained successfully.

\begin{figure}[!h]
\begin{center}
\includegraphics[width=10cm]{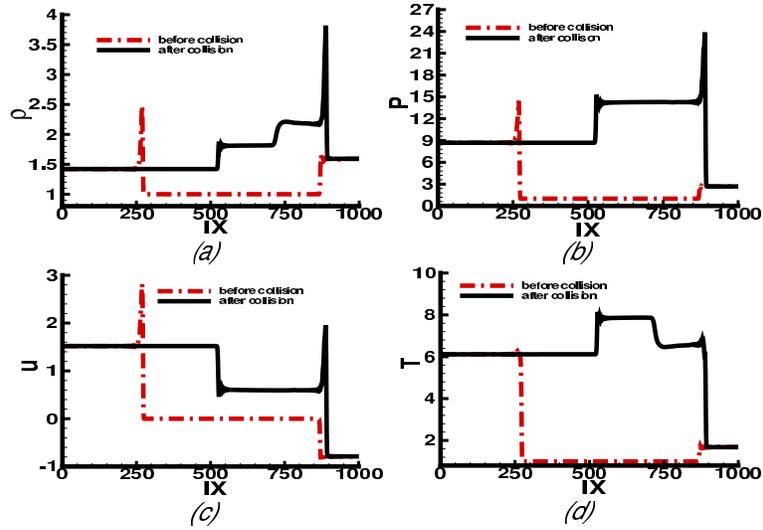}
\caption{\label{collision2} Physical quantity profiles for collision of detonation and shock waves.  Here $M_s=1.5$. The red dash-dotted and solid lines indicate the physical quantities before and after collision, respectively. Figures (a)-(d) are for the  density $\rho$, pressure $P$, $x$-component of velocity $u$ and temperature $T$, respectively.}
\end{center}
\end{figure}

The physical quantities, such as density $\rho$, velocity $u$, pressure $P$, and temperature $T$ before and after collision for detonation and shock with $M_s=1.5$ are given in figure \ref{collision2}(a)-(d). We can find that the density and temperature change after collision while the pressure and velocity remain constants. Before collision, the gas velocity after shock (the direction is to the left) is small than the gas velocity after detonation wave (the direction is to the right). After collision, the direction of gas velocity is to the right. Under the current parameters the density on the right of the contact surface is greater than that on the left.

From figure \ref{collision}(b) and figure \ref{collision2}, we obtain the velocities of incident detonation wave and incident shock are $5.0$ and $-2.1$, respectively. The velocities of transmitted detonation and transmitted shock waves are $4.5$ and $-2.7$ under current parameters. Now, we numerically validate our model via two sets of Hugoniot relations. One set is for the transmitted detonation wave. The other set is for the transmitted shock wave.
We first apply values of the physical quantities before and after the transmitted detonation wave to \eqref{hugoniot1}-\eqref{hugoniot3}.  The values of the LHS are $8.41906$, $11.6014$, $4.89218$. The values of the RHS are $8.48053$, $11.6567$, $4.92229$. Compared with values of the LHS, the relative differences of the two sides are $0.7\%$, $0.5\%$, and $0.6\%$,  respectively.
Likewise, we apply values of the physical quantities before and after the transmitted shock wave to the Hugoniot relations of shock wave, i.e.  \eqref{hugoniot1}-\eqref{hugoniot3} with $\lambda Q =0$. The values of the LHS are $-5.99507,5.58746,1.75327$ and values of the RHS are $-5.97939,5.53874,1.75295$.
Compared with values of the LHS, the relative differences of the two sides are $0.3\%$, $0.9\%$, and $0.02\%$,  respectively.
It is clear that our numerical results agree well with Hugoniot relations of detonation/shock waves.


\subsection{Regular reflection and Mach reflection}

\begin{figure}[!h]
\begin{center}
\includegraphics[width=6cm]{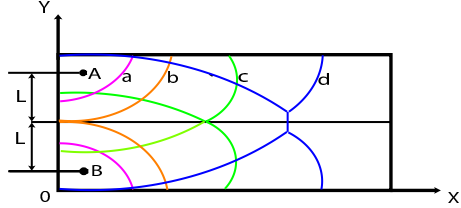}
\caption{\label{sketch} Sketch of regular and Mach reflections of detonation wave.}
\end{center}
\end{figure}

We consider the regular and Mach reflection of detonation wave. Figure \ref{sketch} illustrates a rectangular computational domain where combustible mixture is uniformly congested, two points A and B are measured $2L$. Ignition arises at A and B, then two symmetric detonation waves labeled by $``a"$ generate. Two waves collide at the symmetry plane and obtain the waves labeled by $``b"$. After collision, detonation waves continue to propagate, at the same time, the regular reflection is observed. As the interaction time goes on, the regular reflection changes gradually to the Mach reflection. The initial state is set as blow
\begin{equation} (\rho,u,v,T,\lambda)\mid_{x, y, 0}=
\begin{cases} (1.35826,0.816497,0,2.59709,1) & \text{at~points~A~and~B,} \\
(1,0,0,1,0) & \text{else.}
\end{cases}
\end{equation}
The resolution is set to $\Delta x=\Delta y=0.001$, the time step is set to $\Delta t=10^{-5}$, the relaxation time is $\tau=10^{-5}$, the reaction heat is $Q=1.0$ and the lattice size $N_x\times N_y=400\times400$. The parameters of chemical reaction model are $a=1.0$, $b=10^3$, respectively.

\begin{figure}[!h]
\begin{center}
\includegraphics[width=10cm]{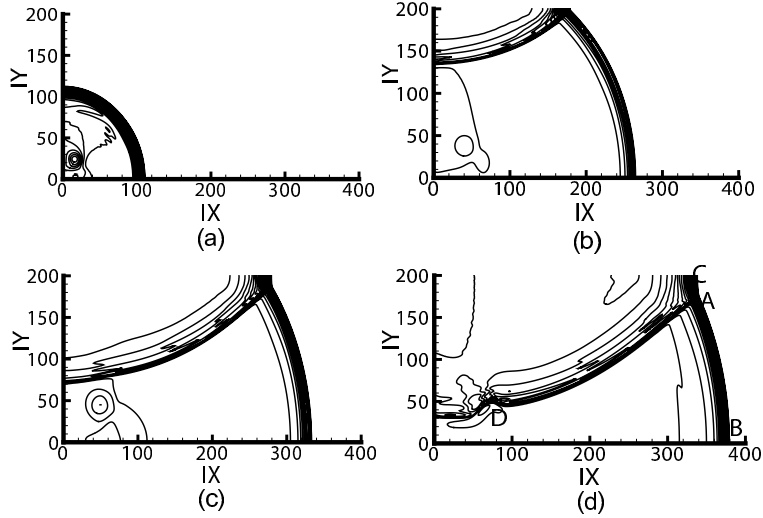}
\caption{\label{regular} The transition process from regular reflection to Mach reflection. (a) Ignition arises and detonation waves generate. (b) The regular reflection of detonation wave. (c) Conversion from the regular reflection to Mach reflection. (d) The Mach reflection of detonation wave.}
\end{center}
\end{figure}

We show the transition process from the regular reflection to Mach reflection in figure \ref{regular}. For simplicity, only one half iso-pressure contour lines at different times are shown. Figure \ref{regular}(a) is for the process of ignition arising and detonation waves generating. Figure \ref{regular}(b) shows the regular reflection. When the incident angle $\alpha_0$ exceeds the critical value $\alpha_{c}$ \cite{courant}, the regular reflection is replaced by the Mach reflection, see figure \ref{regular}(c). The critical value satisfies following equation
 \begin{equation}
\label{angle}
M^2(1-\mu ^2)^2(t_0-t_2)+M\{(1-\mu ^2)^2-(t_0-t_2)^2-(\mu ^2+t_0 t_2)^2\}-(t_0-t_2)=0,
\end{equation}
where $M$ is the Mach number, $t_0=\tan\alpha_{c}$, $\mu ^2=\frac{\gamma-1}{\gamma+1}$, $t_2$ is the unique real root of equation \eqref{angle}. The analytical value of critical angle in this case is $\alpha_{c}=27.668^\circ$. The critical angle from LB result is $\alpha_{c}=26.4323^\circ$. Compared with the analytical solution, the relative difference is about $4\%$.
Figure \ref{regular}(d) is for the Mach reflection of detonation wave. Triple point $A$ appears near the interaction point of detonation wave. In figure \ref{regular}(d) the front $AB$ shows the incident detonation wave, the front $AC$ shows the Mach rod, and $AD$ denotes the reflected shock wave.

\subsection{Richtmyer-Meshkov instability by detonation wave}

Two main types of Richtmyer-Meshkov (RM) instability problems caused by detonation wave are discussed. The first occurs when a detonation wave travels from a light medium to a heavy one. The second occurs when the detonation wave travels from a heavy medium to a light one.

\begin{figure}[!h]
\begin{center}
\includegraphics[width=10cm]{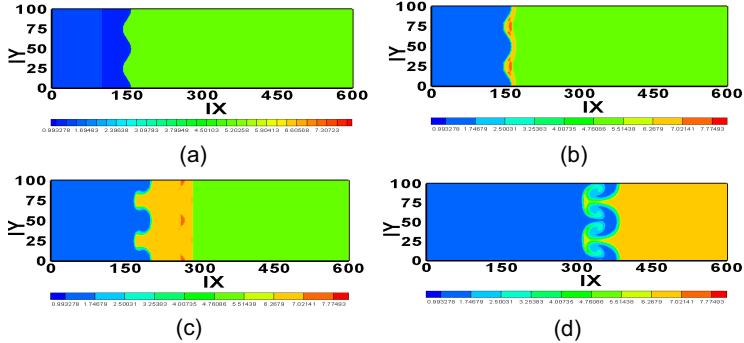}
\caption{\label{rm1} Snapshots of density field for the RM instability in the case where detonation wave travels from light to heavy media.  (a) $t=0$, (b) $t=0.06$, (c) $t=0.2$, (d) $t=1.0$.}
\end{center}
\end{figure}

\emph{Case(I) Detonation wave travels from light to heavy media.} An incident detonation is a strong shock wave propagating into a reactant from the left, followed by a thin zone of reaction which supports the shock. The reactant is heated by the shock via compression, so  that the ignition arises, then detonation hits an interface with sinusoidal perturbation. The initial macroscopic quantities are set as below
\begin{subequations}
\label{case1}
\begin{numcases}{}
(\rho,u,v,T,\lambda)_L=(1.25581,0.34570,0,1.26346,0),\\[5mm]
(\rho,u,v,T,\lambda)_M=(1,0,0,1,0),\\[5mm]
(\rho,,v,T,\lambda)_R=(5.04,0,0,0.198413,0).
\end{numcases}
\end{subequations}
The resolution is set to $\Delta x=\Delta y=0.001$, the time step is $\Delta t=10^{-5}$, the relaxation time is $\tau=10^{-5}$ and lattice size is $N_x\times N_y=600\times100$, $\gamma=2.0$, $a=1.0$, and $b=10^3$ in the whole computational domain. The initial sinusoidal perturbation at the interface is $x=0.15\times N_x\times \Delta x+0.008\times \cos(40\pi y)$. At the open ends, i.e., at the left and right boundaries, we impose the in flow and outflow boundary conditions. The upper and lower boundaries are periodic. Figure \ref{rm1} shows the snapshots of density field, figures (a)-(d) correspond to $t=0$, $0.06$, $0.2$, and $1.0$, respectively. Mushroom structures for detonation wave travel from heavy medium to light one and vesicular structures for detonation travel from light medium to heavy one are successfully obtained, which is similar to the case with shock wave \cite{vis-chen}.

\emph{Case(II) Detonation wave travels from heavy to light media.} Consider a detonation with Mach number 2.5 impinges on a sinusoidal perturbation $x=0.15\times N_x\times \Delta x+0.008\times \cos(40\pi y)$. The initial condition are set to
\begin{subequations}
\label{case2}
\begin{numcases}{}
(\rho,u,v,T,\lambda)_L=(2.27273,1.9799,0,3.52,0),\\[5mm]
(\rho,u,v,T,\lambda)_M=(1,0,0,1,0),\\[5mm]
(\rho,u,v,T,\lambda)_R=(0.33333,0,0,3,0).
\end{numcases}
\end{subequations}

\begin{figure}[!h]
\begin{center}
\includegraphics[width=10cm]{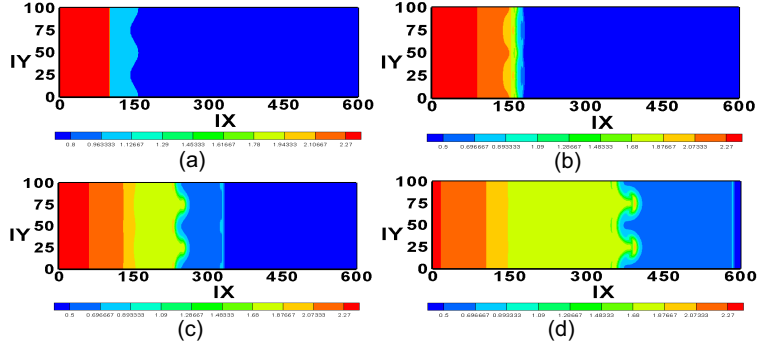}
\caption{\label{rm2} Snapshots of density field for the RM instability in the case where detonation wave travels from heavy to light media.  (a) $t=0$, (b) $t=0.02$, (c) $t=0.05$, (d) $t=0.1$.}
\end{center}
\end{figure}

In the simulation, we adopt the same boundary condition as in case(I). The computational domain is $0.6\times 0.1$, the common parameters are $\Delta x=\Delta y=0.001$, $\Delta t=10^{-5}$, $\tau=10^{-5}$, $\gamma=2.0$, $a=1.0$, and $b=10^3$. Figure \ref{rm2} shows snapshots of density field, figure (a)-(d) correspond to the time $t=0$, $0.02$, $0.05$, and $0.1$ respectively. In this case, interface reversal is successfully observed.

\section{LB study on detonation phenomena}

\subsection{Phase diagram of viscous detonation}

Presently, most detonation numerical studies reported are based on solving governing equations which couples Euler equation and chemical kinetics equation. This method neglects the effect of transport processes. The LB scheme, which naturally includes the effects of viscosity and heat conduction, describes more faithfully real physical procedure than traditional method.

\begin{figure}[!h]
\begin{center}
\includegraphics[width=10cm]{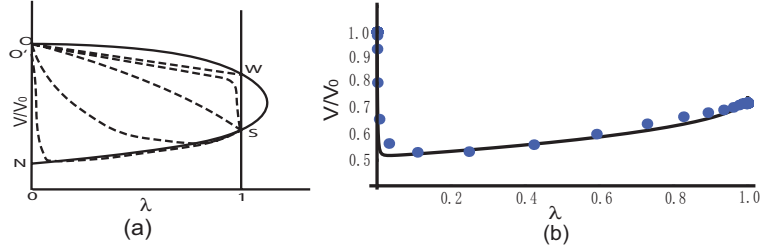}
\caption{\label{viscous} Phase diagram of viscous detonation. (a) Sketch of phase diagram. (b) Numerical results by Mathematica 8.0 and our LB scheme.}
\end{center}
\end{figure}

Consider the one-dimensional steady flow of detonation \cite{fickett}. Wood has shown the $\lambda$-$V/V_0$ phase plane for detonation \cite{wood}, a sketch is shown in figure \ref{viscous}(a). The vertical lines $\lambda=0$ and $\lambda=1$ indicate Hugoniot curves for shock and detonation, respectively. The solid line is Rayleigh curve, which intersects the vertical line $\lambda=1$ at points S and W, corresponding to the strong detonation and weak detonation respectively. The dashed lines are integral curves for detonation depending on different values of detonation velocity D. In this plane, the state point in the ZND model jumps discontinuously from initial point O to von Neumann point N, then moves up the Rayleigh curve to point $S$. The analogous solution of the viscous problems is an integral curve leaving point $O$ or $O'$ and terminating at point S, without passing through von Neumann point N.

\begin{figure}[!h]
\begin{center}
\includegraphics[width=10cm]{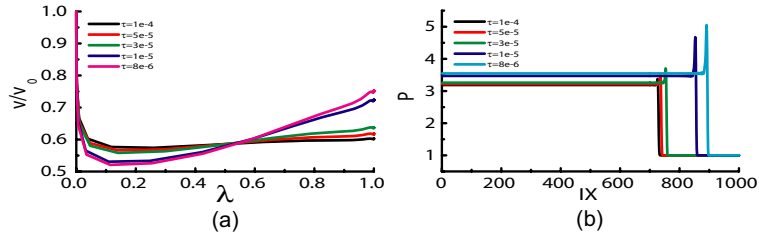}
\caption{\label{lbvisco} The influences of viscosity and heat conductivity. (a) Influence on the $\lambda$-$V/V_0$ phase diagram. (b) Influence on the von Neumann peak.}
\end{center}
\end{figure}

The $\lambda$-$V/V_0$ phase diagram for the viscous detonation obtained by our LB model is showed in Figure \ref{viscous}(b), see the solid circles, where the line is for the integral curve solved by Mathematica 8.0. The parameters used here are the same as those in figure \ref{piston}.
The deviation of the integral curve from LB result originates from the different values of viscosity used in the integration of procedure by Mathematica 8.0. For Mathematica 8.0, the viscosity is $P_{cj}\tau$,
while it equals $P\tau$ and is variable in our LB simulating. Figure \ref{lbvisco}(a) and figure \ref{lbvisco}(b) show the influences of viscosity and heat conductivity. As the relaxation time $\tau$ increases, the curve for $V/V_0$ versus $\lambda$ become flatter, and the von Neumann peaks becomes less evident, the viscosity smoothes wave front of pressure.

\subsection{Thermodynamic equilibrium versus chemical reaction}

Among the seven moment relations \eqref{mo1} to \eqref{mo7} required by the present LB model, only for the first three, i.e. the definitions of density, momentum and energy (see equations \eqref{mo1} to \eqref{mo3}), the equilibrium distribution function $f^{eq}_{ki}$ can be replaced by the distribution function $f_{ki}$. If we replace $f^{eq}_{ki}$ by $f_{ki}$ in the left hand side of any one of equations \eqref{mo4} to \eqref{mo7}, the value of left hand side will have a deviation from that of the right hand side. This deviation may work as a measure for the deviation of system from its thermodynamic equilibrium \cite{front}. We introduce
\begin{equation}
\boldsymbol{\Delta}_m=\textbf{I}_m(f_{ki})-\textbf{I}_m(f_{ki}^{eq}),~~~~~(m=4,5,6,7)
\label{mom}
\end{equation}
where the subscript $m$ indicate the $m$-th moment, $\textbf{I}_m(f_{ki})$ and $\textbf{I}_m(f_{ki}^{eq})$ are definitions of moments calculate by $f_{ki}$ and $f_{ki}^{eq}$, respectively.
Specifically,
\begin{equation}
\Delta_{4,\alpha\beta}=I_{4,\alpha\beta}(f_{ki})-I_{4,\alpha\beta}(f_{ki}^{eq})=\sum_{ki}v_{ki\alpha}v_{ki\beta}f_{ki}
-\sum_{ki}v_{ki\alpha}v_{ki\beta}f^{eq}_{ki},
\label{mom4}
\end{equation}
where $\Delta_{4,\alpha\beta}$, $I_{4,\alpha\beta}(f_{ki})$ and $I_{4,\alpha\beta}(f_{ki}^{eq})$ are the components $xx$, $xy$, $yx$, $yy$, respectively.
\begin{equation}
\Delta_{5,\alpha\beta\gamma}=I_{5,\alpha\beta\gamma}(f_{ki})-I_{5,\alpha\beta\gamma}(f_{ki}^{eq})=\sum_{ki}v_{ki\alpha}v_{ki\beta}v_{ki\gamma}f_{ki}
-\sum_{ki}v_{ki\alpha}v_{ki\beta}v_{ki\gamma}f^{eq}_{ki},
\label{mom5}
\end{equation}
where $\Delta_{5,\alpha\beta\gamma}$, $I_{5,\alpha\beta\gamma}(f_{ki})$ and $I_{5,\alpha\beta\gamma}(f_{ki}^{eq})$ are components
 $xxx$, $xxy$, $xyx$, $xyy$, $yxx$, $yxy$, $yyx$, $yyy$, respectively.
\begin{equation}
\Delta_{6,\beta}=I_{6,\beta}(f_{ki})-I_{6,\beta}(f_{ki}^{eq})=\sum_{ki}\frac{1}{2}v_k^2v_{ki\alpha}f_{ki}
-\sum_{ki}\frac{1}{2}v_k^2v_{ki\alpha}f^{eq}_{ki},
\label{mom6}
\end{equation}
where $\Delta_{6,\beta}$, $I_{6,\beta}(f_{ki})$ and $I_{6,\beta}(f_{ki}^{eq})$ are the $x$- and $y$-components, respectively.
\begin{equation}
\Delta_{7,\alpha\beta}=I_{7,\alpha\beta}(f_{ki})-I_{7,\alpha\beta}(f_{ki}^{eq})=\sum_{ki}\frac{1}{2}v_k^2v_{ki\alpha}v_{ki\beta}f_{ki}
-\sum_{ki}\frac{1}{2}v_k^2v_{ki\alpha}v_{ki\beta}f^{eq}_{ki},
\label{mom7}
\end{equation}
where $\Delta_{7,\alpha\beta}$, $I_{7,\alpha\beta}(f_{ki})$ and $I_{7,\alpha\beta}(f_{ki}^{eq})$ are the components $xx$, $xy$, $yx$, $yy$, respectively.
In summary, among the four moments $\textbf{I}_4$ to $\textbf{I}_7$, $\textbf{I}_4$ and $\textbf{I}_7$ are second-order tensors, $\textbf{I}_5$ is a third-order tensor and $\textbf{I}_6$ is a first-order tensor or vector.

It is clear that $\boldsymbol{\Delta}_m$ and $\textbf{I}_m$ contains the information of the macroscopic flow velocity $\textbf{u}$ Furthermore, we replace $\textbf{v}_{ki}$ by $\textbf{v}_{ki}-\textbf{u}$ in any one of equations \eqref{mom4} to \eqref{mom7}, named $\boldsymbol{\Delta}_m^*$ and $\textbf{I}_m^*$. $\boldsymbol{\Delta}_m^*$ and $\textbf{I}_m^*$ are only the manifestation of the thermo-fluctuations of molecules relative to the macroscopic flow velocity $\textbf{u}$. Specifically,
\begin{eqnarray}
\label{mom41}&~& \Delta_{4,\alpha\beta}^*=I_{4,\alpha\beta}^*(f_{ki})-I_{4,\alpha\beta}^*(f_{ki}^{eq})=\sum_{ki}(f_{ki}-f^{eq}_{ki})
(v_{ki\alpha}-u_\alpha)(v_{ki\beta}-u_\beta),\\
\label{mom51}&~& \Delta_{5,\alpha\beta\gamma}^*=I_{5,\alpha\beta\gamma}^*(f_{ki})-I_{5,\alpha\beta\gamma}^*(f_{ki}^{eq})=\sum_{ki}(f_{ki}-f^{eq}_{ki})
(v_{ki\alpha}-u_\alpha)(v_{ki\beta}-u_\beta)(v_{ki\gamma}-u_\gamma),\\
\label{mom61}&~& \Delta_{6,\beta}^*=I_{6,\beta}^*(f_{ki})-I_{6,\beta}^*(f_{ki}^{eq})=\sum_{ki}\frac{1}{2}(f_{ki}-f^{eq}_{ki})
(v_{ki\alpha}-u_\alpha)^2(v_{ki\beta}-u_\beta),\\
\label{mom71}&~& \Delta_{7,\alpha\beta}^*=I_{7,\alpha\beta}^*(f_{ki})-I_{7,\alpha\beta}^*(f_{ki}^{eq})
=\sum_{ki}\frac{1}{2}(f_{ki}-f^{eq}_{ki})
(v_{ki\alpha}-u_\alpha)^2(v_{ki\beta}-u_\beta)(v_{ki\gamma}-u_\gamma).
\end{eqnarray}
where, $\boldsymbol{\Delta}_m^*$ and $\textbf{I}_m^*$ have the same property components as $\boldsymbol{\Delta}_m$ and $\textbf{I}_m$ respectively.

\begin{figure}[!h]
\begin{center}
\includegraphics[width=10cm]{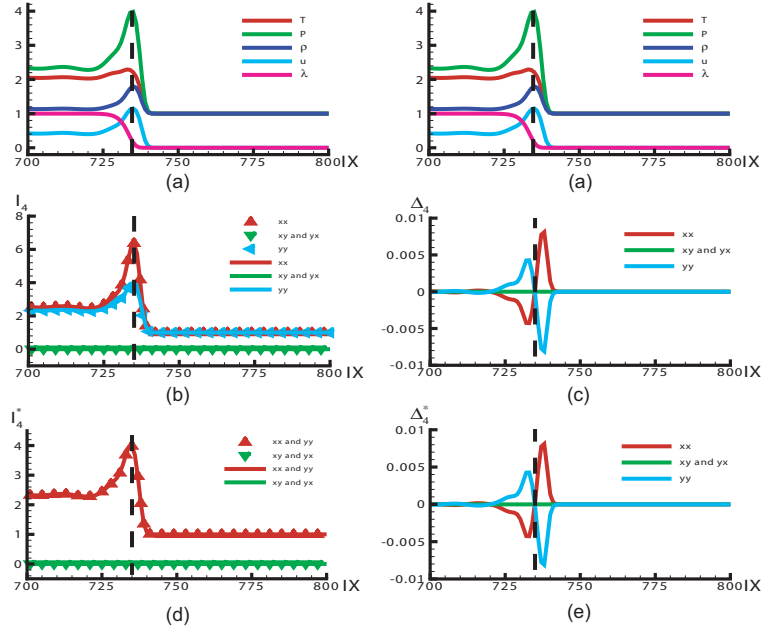}
\caption{\label{moment4} Profiles of physical quantities. Figure (a) is for the density $\rho$, pressure $P$, temperature $T$, $x$-component of velocity $u$ and the fraction of product $\lambda$. Figures (b) is for the moment $\textbf{I}_4$. Figures (c) is for the deviation $\boldsymbol{\Delta}_4$. Figures (d) is for the moment $\textbf{I}_4^*$. Figures (e) is for the deviation $\boldsymbol{\Delta}_4^*$. In figures (b) and (d) the symbols are for moments calculated from $f_{ki}$ and the lines are for moments calculated from $f_{ki}^{eq}$. All components of $\textbf{I}_4$, $\textbf{I}_4^*$, $\boldsymbol{\Delta}_4$ and $\boldsymbol{\Delta}_4^*$ are shown. The specific correspondences are referred to the legends. To shown clearly the system state from various sides, Figure (a) is shown two times in the two columns to guide the eyes. The vertical dashed line in each plot indicates the von Neumann peak.}
\end{center}
\end{figure}

\begin{figure}[!h]
\begin{center}
\includegraphics[width=10cm]{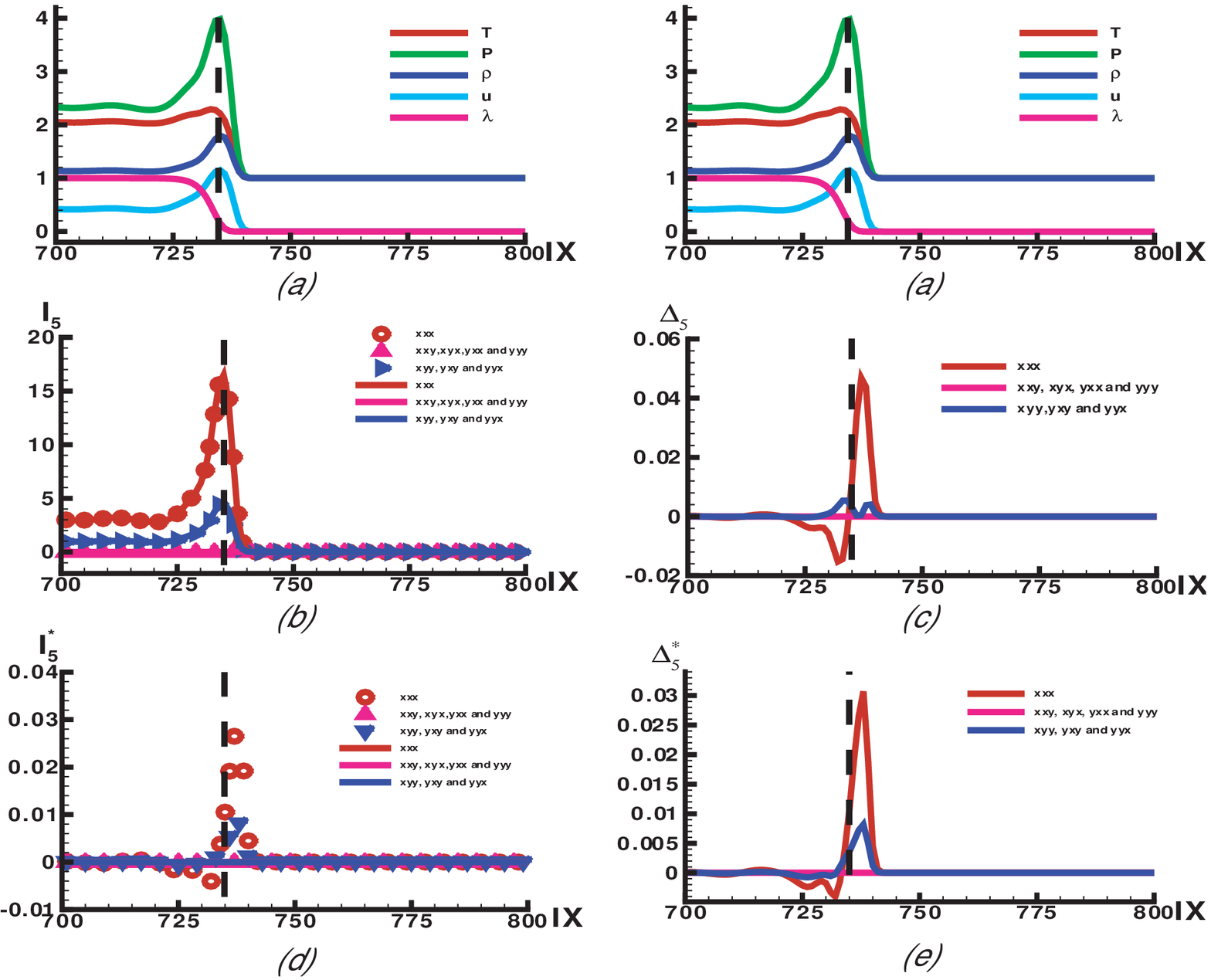}
\caption{\label{moment5} Profiles of physical quantities. Figure (a) is for the density $\rho$, pressure $P$, temperature $T$, $x$-component of velocity $u$ and the fraction of product $\lambda$. Figures (b) is for the moment $\textbf{I}_5$. Figures (c) is for the deviations $\boldsymbol{\Delta}_5$. Figures (d) is for the moment $\textbf{I}_5^*$. Figures (e) is for the deviations $\boldsymbol{\Delta}_5^*$. In figures (b) and (d) the symbols are for moments calculated from $f_{ki}$ and the lines are for moments calculated from $f_{ki}^{eq}$. All components of $\textbf{I}_5$, $\textbf{I}_5^*$, $\boldsymbol{\Delta}_5$ and $\boldsymbol{\Delta}_5^*$ are shown. The specific correspondences are referred to the legends. To shown clearly the system state from various sides, Figure (a) is shown two times in the two columns to guide the eyes.  The vertical dashed line in each plot indicates the von Neumann peak.}
\end{center}
\end{figure}

\begin{figure}[!h]
\begin{center}
\includegraphics[width=10cm]{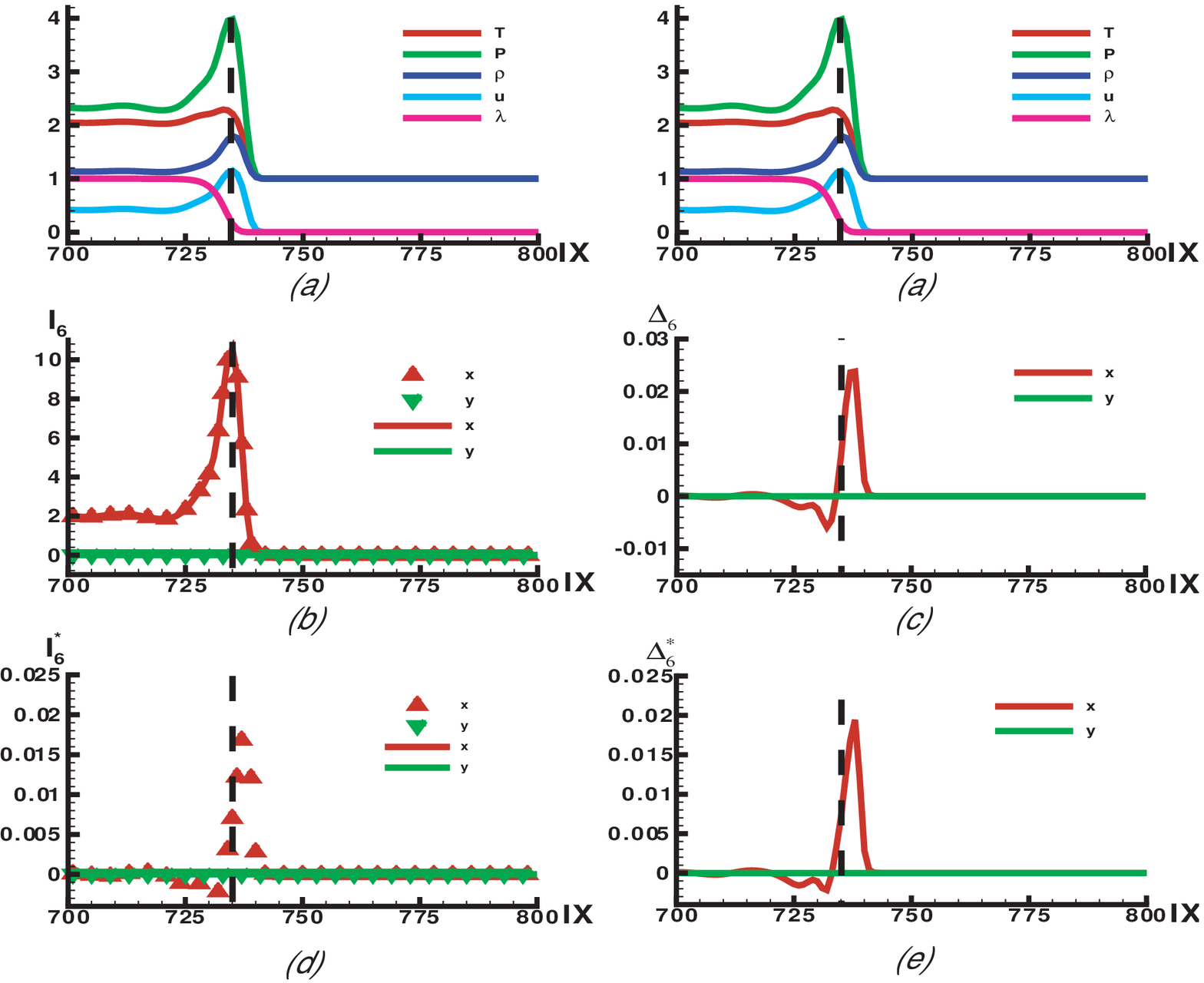}
\caption{\label{moment6} Profiles of physical quantities. Figure (a) is for the density $\rho$, pressure $P$, temperature $T$, $x$-component of velocity $u$ and the fraction of product $\lambda$. Figures (b) is for the moment $\textbf{I}_6$. Figures (c) is for the deviations $\boldsymbol{\Delta}_6$. Figures (d) is for the pure thermodynamic moment $\textbf{I}_6^*$. Figures (e) is for the pure thermodynamic
deviations $\boldsymbol{\Delta}_6^*$. In figures (b) and (d) the symbols are for moments calculated from $f_{ki}$ and the lines are for moments calculated from $f_{ki}^{eq}$. All components of $\textbf{I}_6$, $\textbf{I}_6^*$, $\boldsymbol{\Delta}_6$ and $\boldsymbol{\Delta}_6^*$ are shown. The specific correspondences are referred to the legends. To shown clearly the system state from various sides, Figure (a) is shown two times in the two columns to guide the eyes.  The vertical dashed line in each plot indicates the von Neumann peak.}
\end{center}
\end{figure}

\begin{figure}[!h]
\begin{center}
\includegraphics[width=10cm]{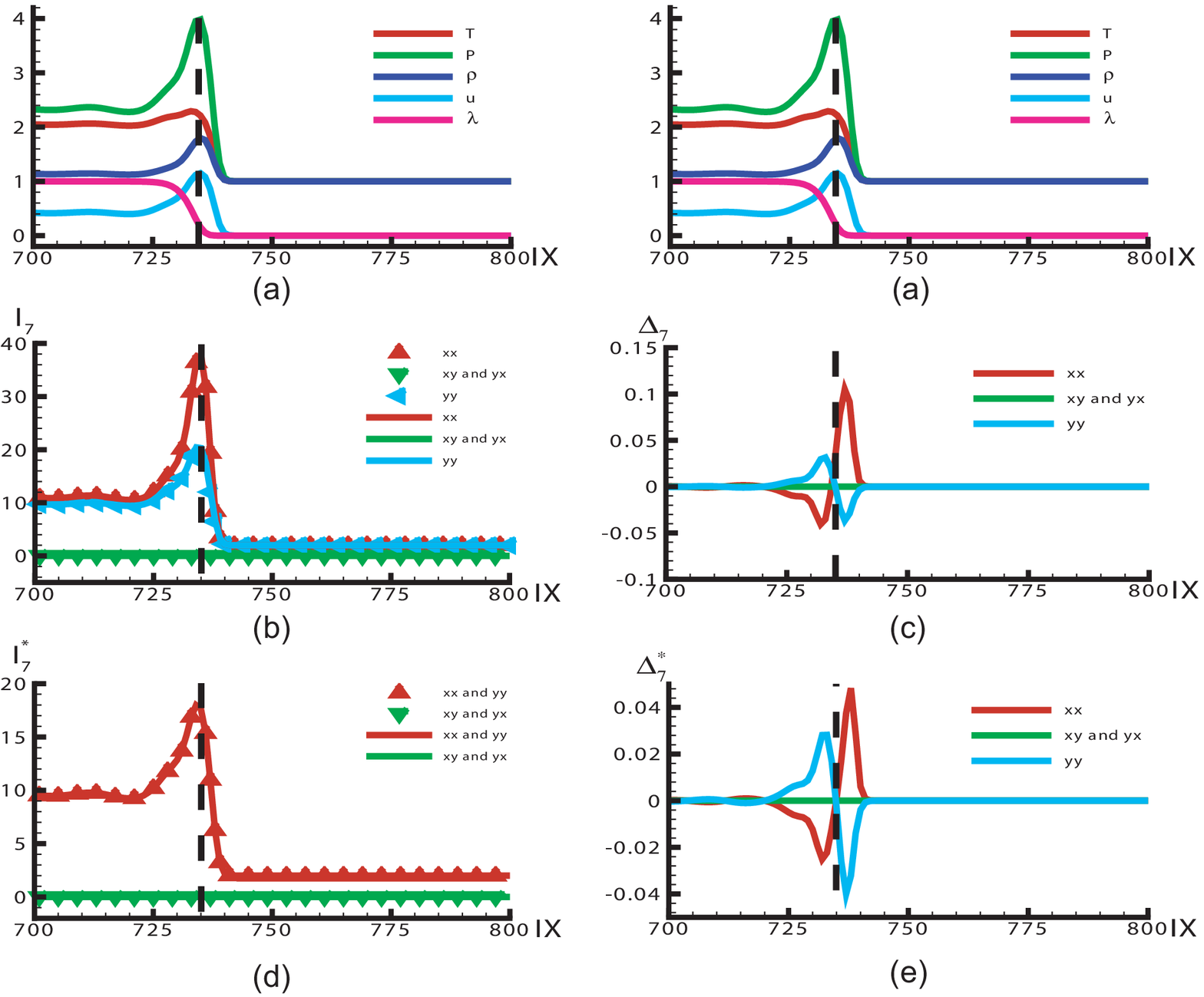}
\caption{\label{moment7} Profiles of physical quantities. Figure (a) is for the density $\rho$, pressure $P$, temperature $T$, $x$-component of velocity $u$ and the fraction of product $\lambda$. Figures (b) is for the moment $\textbf{I}_7$. Figures (c) is for the deviations $\boldsymbol{\Delta} _7$. Figures (d) is for the pure thermodynamic moment $\textbf{I}_7^*$. Figures (e) is for the pure thermodynamic
deviations $\boldsymbol{\Delta}_7^*$. In figures (b) and (d) the symbols are for moments calculated from $f_{ki}$ and the lines are for moments calculated from $f_{ki}^{eq}$. All components of $\textbf{I}_7$, $\textbf{I}_7^*$, $\boldsymbol{\Delta}_7$ and $\boldsymbol{\Delta}_7^*$ are shown. The specific correspondences are referred to the legends. To shown clearly the system state from various sides, Figure (a) is shown two times in the two columns to guide the eyes.  The vertical dashed line in each plot indicates the von Neumann peak.}
\end{center}
\end{figure}

\begin{figure}[!h]
\begin{center}
\includegraphics[width=10cm]{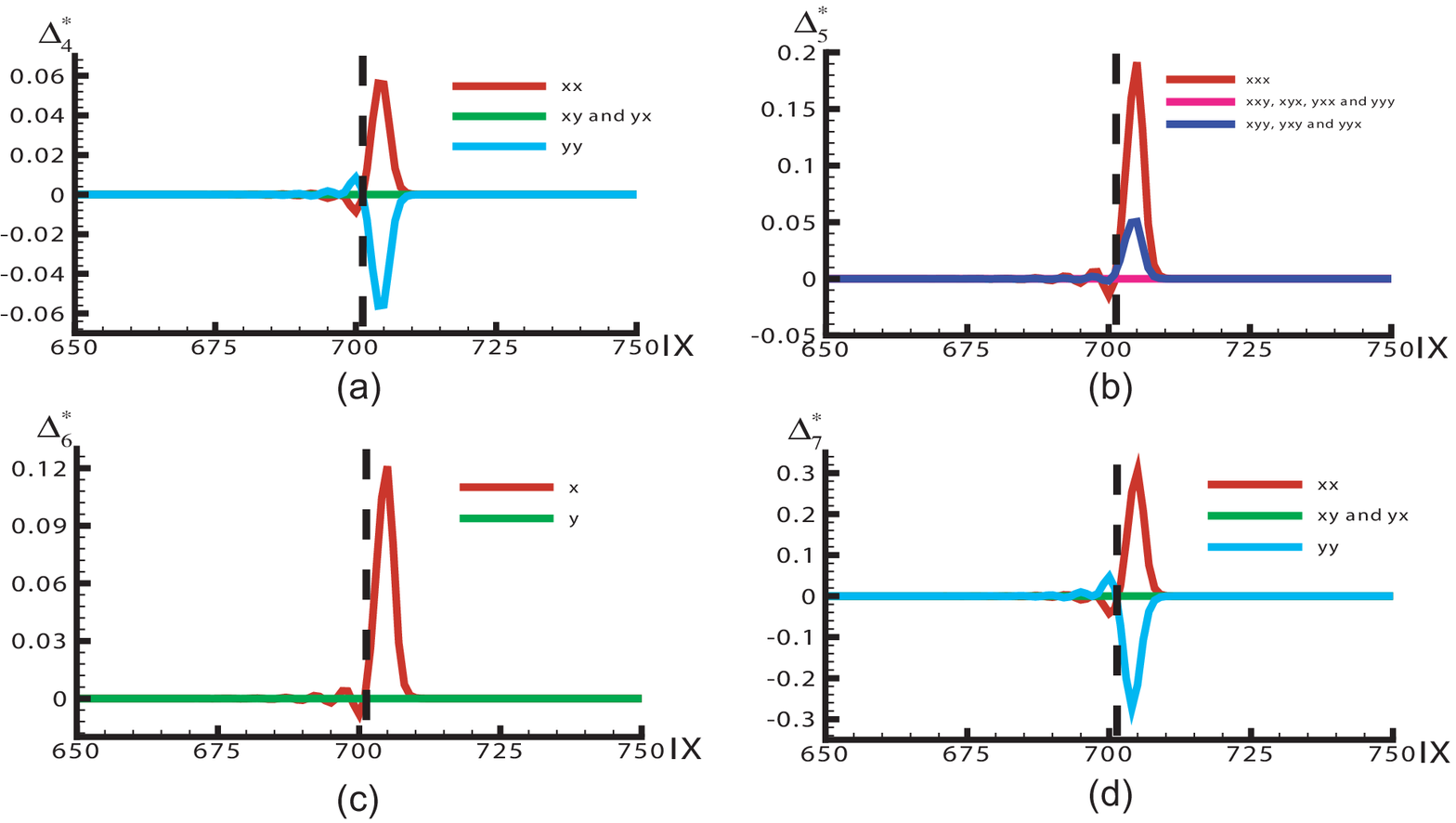}
\caption{\label{pure} The deviations $\boldsymbol{\Delta}_4^*$ to $\boldsymbol{\Delta}_7^*$ with relaxation factor $\tau=10^{-4}$. Figures (a)-(d) are for $\boldsymbol{\Delta}_4^*$ to $\boldsymbol{\Delta}_7^*$, respectively.}
\end{center}
\end{figure}

Now, we use the newly introduced concepts and theory to study a simple case of detonation. For a similar case of CJ detonation as shown in figure \ref{piston}, figures \ref{moment4} to \ref{moment7} show the fourth moments and its deviations to the seventh moments and its deviations respectively. Figure \ref{moment4}(a) to \ref{moment7}(a) show the profiles of physical quantities: the density $\rho$, pressure $P$, temperature $T$, $x$-component of velocity $u$ and the fraction of product $\lambda$. The initial macroscopic quantities are set as below
\begin{subequations}
\label{pistoncase21}
\begin{numcases}{}
(\rho,u,v,T,\lambda)_L=(1.20332,0.535113,0,2.17107,1),\\[5mm]
(\rho,u,v,T,\lambda)_R=(1,0,0,1,0).
\end{numcases}
\end{subequations}
The relaxation time is $\tau=10^{-5}$. Other parameters are chosen as the same as figure \ref{piston}. Figure \ref{moment4}(b) to \ref{moment7}(b) are for $\textbf{I}_4$ to $\textbf{I}_7$ respectively. Figure \ref{moment4}(c) to \ref{moment7}(c) are for $\boldsymbol{\Delta}_4$ to $\boldsymbol{\Delta}_7$ respectively. Figure \ref{moment4}(d) to \ref{moment7}(d) are for $\textbf{I}_4^*$ to $\textbf{I}_7^*$  respectively. Figure \ref{moment4}(e) to \ref{moment7}(e) are for $\boldsymbol{\Delta}_4^*$ to $\boldsymbol{\Delta}_7^*$ respectively. In figure \ref{moment4}(b) to \ref{moment7}(b) and \ref{moment4}(d) to \ref{moment7}(d), the symbols are for moments calculated from $f_{ki}$ and the solid lines are for moments calculated from $f_{ki}^{eq}$.  The specific correspondences between the components and the symbols/lines are referred to the legends. To shown clearly the system state from various sides, in figures \ref{moment4}(a) to \ref{moment7}(a), the profiles of $\rho$, $P$, $T$, $u$ and $\lambda$ are repeatedly shown two times in the two columns to guide the eyes. The vertical dashed line in each plot indicates the von Neumann peak. Here the time $t=0.3$. Figure \ref{pure} shows the profiles of $\boldsymbol{\Delta}_4^*$ to $\boldsymbol{\Delta}_7^*$ with relaxation factor $\tau=10^{-4}$ at time $t=0.3$.

From figure \ref{moment4}(a) we can find that, due to the compression effects of the shock wave, the density, pressure, temperature and flow velocity increase sharply and reach the von Neumann peak. When the temperature arrives at the threshold value for explosion, chemical reaction occurs. The fraction of product $\lambda$ increases from $0$ to $1$ within the reaction zone. Since the propagation of detonation wave is faster than the pure shock wave, compared with the case of pure shock wave action, the matter behind the front of detonation wave expand quickly within the reaction time. Therefore, the density, pressure and flow velocity decrease quickly within the reaction zone. As for the  variation of temperature, the situation is a little more complex. The volume expansion results in a decrease of temperature, while the chemical reaction releases heat to increase the temperature. Finally, the variation of temperature is dependent on the competition of the two mechanisms. More specifically, it is dependent on the equation of state and the reaction model. It should be pointed out, in most of the traditional studies which are based on the Euler equations, the width of the shock front is assumed to be zero. So, the density, pressure, temperature and flow velocity reach their von Neumann peak values instantaneously. Recent studies have shown that this assumption is not strictly true. Our simulation results clearly show the increasing processes of these quantities. For the case shown in figure \ref{piston} or figure \ref{moment4}(a), we clearly observe that the chemical reaction occurs before the temperature arrives at its von Neumann peak. This result can also be understood from figure \ref{viscous}. If the threshold temperature increases, the starting point of the chemical reaction will be more closer to the von Neumann peak. If the threshold temperature is higher than the value at the von Neumann peak, no chemical reaction occurs. What we observe will be a pure shocking process. The density, pressure, temperature and flow velocity will keep constant after the front of the shock wave.

From figures \ref{moment4} it is clear that the moments calculated from $f_{ki}$ behavior qualitatively the same as those from $f_{ki}^{eq}$. The $\textbf{I}_4$ and $\textbf{I}_4^*$ have components showing a peak at the same position as the von Neumann peak and have components being nearly zero. The $xx$ and $yy$ components of $\textbf{I}_4^*$ have the same amplitude. The $\boldsymbol{\Delta}_4$ and $\boldsymbol{\Delta}_4^*$, have components deviating significantly from zero during the reaction procedure. It is very interesting to find that, at the von Neumann peak, the system is much closer to its thermodynamic equilibrium. In the front of and behind the von Neumann peak, $xx$ and $yy$ components of system deviates from its equilibrium in opposite directions with the same deviation amplitude. It should also be pointed out that the deviation from thermodynamic equilibrium starts from the beginning of the shocking procedure, instead of the beginning of the chemical reaction. With finishing of the chemical reaction, the system goes back to its thermodynamic equilibrium gradually. The features of $\textbf{I}_7$, $\textbf{I}_7^*$, $\boldsymbol{\Delta}_7$ and  $\boldsymbol{\Delta}_7^*$ shown in figure \ref{moment7} are similar.
Figure \ref{moment5} shows that the $\textbf{I}_5$ and $\textbf{I}_5^*$ have components showing a peak at the same position near the von Neumann peak and have components being nearly zero. At the von Neumann peak, the system is near its thermodynamic equilibrium.
Since $\boldsymbol{\Delta}_6$ and $\boldsymbol{\Delta}_6^*$ shown in figure \ref{moment6} has only two components, the amplitude of its $y$ component keeps zero, only the $x$ component shows the deviation from thermodynamic equilibrium.
Figure \ref{pure} shows that, when $\tau$ increases to $10^{-4}$, at the von Neumann peak, the system is closer to its thermodynamic equilibrium, while in front of the peak, the amplitude of deviating from equilibrium becomes much larger.

All the non-equilibrium effects in figure \ref{moment4} to \ref{moment7} can be consistently interpreted as below. Among the four physical fields for the density, momentum, pressure and temperature, the temperature gradient is the most fundamental driving force triggering the non-equilibrium effects. The gradient of any other triggers the non-equilibrium effects via triggering macroscopic transportation which leads to temperature gradient. The temperature gradient first initiates variance of the internal energy in the same degree of the freedom as that of the temperature gradient. Then, part of internal energy variance is transferred to other degrees of freedoms via collisions of molecules. Then, the internal energy in this degree of freedom further varies according to the temperature gradient, and so on. Only when the temperature gradient vanishes, the system can arrives at its thermodynamic equilibrium.


\section{Conclusions and discussions}

The combustion and detonation phenomena are widely used in the acceleration of various projectiles, mining technologies, depositing of coating to a surface or cleaning of equipment, etc. However, unintentional detonation phenomenon, when deflagration is desired, is a problem in some devices. The combustion and detonation phenomena have become highly concerned issues in science and engineering.
 In this work, a LB model for combustion and detonation has been presented.
 This model is composed of descriptions of two processes, the fluid flow and the chemical reaction. The former is described by the FDLB model \cite{phase-gan1}, which gives the same results as the Navier-Stokes equations in the hydrodynamic limit. The chemical reaction is described by the Lee-Tarver model \cite{lee-tarver}. In this scheme, the heat of reaction is coupled with the internal energy. The increment of internal energy results in the variation of pressure which naturally influences the flow behavior. Since the time scale in the chemical reaction process is much smaller than that in the thermodynamical process, an operator-splitting scheme is necessary to obtain a successful simulation.
 In order to indicate the validity of the new model, several problems are studied: (i) the piston problem with viscosity and heat conduction which is a typical test in studies on shock initiation of explosive, (ii) collision between detonation and shock waves, (iii) regular and Mach reflection of plane detonation wave, (iv) the Richtmyer-Meshkov instability, (v) the phase diagram for viscous detonation. Simulation results fully indicate that the present model works for fundamental processes and problems of shock initiation of explosives, can capture the essential features of combustion and detonation, such as the von Neumann peak, chemical reaction zone, and etc.

 In contrast with Yamatomo's model \cite{yamamoto}, the scheme describes the density, momentum and energy using only one distribution function. Compared with the model by Filippova et al \cite{filippova} and Lee et al \cite{lee}, the present model is completely thermal and compressible. Compared with all the previous LB models for combustion \cite{succi-comb,filippova,yamamoto,lee}, the present model realizes nature coupling of the chemical reaction and flow behavior. It works not only for the  combustion systems with low-Mach number but also for the detonation phenomena with high-Mach number.
It can be further used to investigate the fundamental physics in various phenomena, particularly those non-equilibrium effects, related to combustion and detonation. For example, problems on the triple point trajectory, the Mach reflection and regular reflection of detonation waves, problems on shock-to-detonation and deflagration-to-detonation transitions, and problems on the Rayleigh-Taylor, Richtmyer-Meshkov and Kelvin-Helmholtz instabilities induced by detonation wave, etc. The same ideas, including that for the operator-splitting,  can be used to construct appropriate LB models and simulate various combustion and detonation phenomena via choosing corresponding compressible LB and reaction models.

As a specific application of the new model, we studied the simple steady detonation phenomenon. To show the merit of LB model over the traditional ones, we focus on the reaction zone to study the non-equilibrium effects. It is found that, at the von Neumann peak, the system is near its thermodynamic equilibrium. In the front of and behind the von Neumann peak, the system deviates from its equilibrium in opposite directions.  The deviation from thermodynamic equilibrium starts from the beginning of the shocking procedure, instead of the beginning of the chemical reaction. With finishing of the chemical reaction, the system goes back to its thermodynamic equilibrium gradually. Even though most of the combustion and detonation phenomena show three-dimensional effects, if we observe from a region being small enough so that the wave front can be roughly regarded as a plane, the one-dimensional conclusions still work. From the deviation from thermodynamic equilibrium, $\boldsymbol{\Delta}_m^*$, defined in this paper, we can understand more on the macroscopic effects of the system deviates from its thermodynamic equilibrium. If we assume the normal direction of the wave front is $\mathbf{n}$ and $\mathbf{t}$ is an arbitrarily chosen tangential direction, then we have the following observations.
The $nn$ and the $tt$ components of $\boldsymbol{\Delta}_4^*$ behavior qualitatively in the opposite directions with the same amplitudes. The $nn$ ($tt$) component shows a positive (negative) peak and a negative (positive) peak in the front of and behind the von Neumann peak, respectively.  The other two components of $\boldsymbol{\Delta}_4^*$ keep zero during the whole procedure. The features of $\boldsymbol{\Delta}_7^*$ are similar. As for $\boldsymbol{\Delta}_6^*$ which has only two components, the amplitude of its $t$ component keeps zero, only the $n$ component shows the deviation from thermodynamic equilibrium. Among the eight components of $\boldsymbol{\Delta}_5^*$, the $nnn$ component has the largest amplitude, then the ones for $ttn$, $tnt$ and $ntt$. The others keep zero. With the same method, the complex three-dimensional effects can be carefully studied and they are out of the scope of the present paper. With increasing the viscosity, the amplitude of deviation from thermodynamic equilibrium becomes larger at the two sides of the von Neumann peak, while at the von Neumann peak the system is closer to its thermodynamic equilibrium.

\section*{Acknowledgements}
The authors would like to sincerely thank Prof. Guoxi Ni, Drs. Yinfeng Dong, Yanbiao Gan, Feng Chen, Chuandong Lin, Weiwei Pang, and Xihua Xu for helpful discussions. AX and GZ acknowledge support of the Science Foundation of China Academy of Engineering Physics [under Grant Nos. 2012B0101014 and 2011A0201002], National Natural Science Foundation of China [under Grant Nos. 11075021 and 91130020] and the Foundation of State Key Laboratory of Explosion Science and Technology.

\end{document}